\documentclass[aps,prd,twocolumn,groupedaddress,nofootinbib]{revtex4}
\pdfoutput=1  
\usepackage{bm}
\usepackage{graphicx,color}
\usepackage{latexsym,amsmath,amssymb,graphicx,booktabs}
\usepackage{hyperref}
\usepackage{placeins}
\usepackage{subfigure}

\definecolor{MyBlue}{rgb}{0.15,0.15,0.70}
\definecolor{Dgreen}{rgb}{0,0.7,0.0}

\hypersetup{
colorlinks=true,
citecolor=MyBlue,
linkcolor=black,
urlcolor=MyBlue
}

\usepackage{amssymb}
\usepackage{amsmath}
\usepackage{amsfonts}
\usepackage{upgreek}
\usepackage{latexsym}
\usepackage{appendix}
\usepackage{eufrak}
\usepackage{dsfont}
\usepackage{mathtools}

\usepackage[export]{adjustbox}

\include{mydefs}

\newcommand\spart{\;\raise1.0pt\hbox{/}\hskip-6pt\partial}
\newcommand\spartb{\;\overline{\raise1.0pt\hbox{/}\hskip-6pt\partial}}

\newcommand{\de}{\delta}
\newcommand{\De}{\Delta}

\newcommand{\be}{\begin{equation}}
\newcommand{\ee}{\end{equation}}
\newcommand{\bea}{\begin{eqnarray}}
\newcommand{\eea}{\end{eqnarray}}
\newcommand{\bean}{\begin{eqnarray*}}
\newcommand{\eean}{\end{eqnarray*}}
\newcommand{\beal}{\begin{align}}
\newcommand{\enal}{\end{align}}

\usepackage{ntheorem}

\newtheorem*{theorem-non}{Theorem}

\newcommand{\PP}{\mathcal{P}}

\newcommand{\mathleft}{\@fleqntrue\@mathmargin0pt}
\newcommand{\mathcenter}{\@fleqnfalse}

\definecolor{revisioncolor}{RGB}{22, 158, 230}

\newcommand{\tj}[6]{ \begin{pmatrix}
   #1 & #2 & #3 \\
   #4 & #5 & #6 
\end{pmatrix}}

\graphicspath{ {./Graphs/} }
\begin{document}

\vspace*{2cm}

\title{On the $2^\text{nd}$ feature of the matter two-point function}
\author{Vittorio Tansella}
\affiliation{D\'epartement de Physique Th\'eorique and Center for Astroparticle Physics, Universit\'e de Gen\`eve, 24 quai Ansermet, CH--1211 Gen\`eve 4, Switzerland}
\email{vittorio.tansella@unige.ch}
\vspace{1 em}
\date{\today}

\begin{abstract}
We point out the existence of a second feature in the matter two-point function, besides the acoustic peak, due to the baryon-baryon correlation in the early universe and positioned at twice the distance of the peak. We discuss how the existence of this feature is implied by the well-known heuristic argument that explains the baryon bump in the correlation function. A standard $\chi^2$ analysis to estimate the detection significance of the second feature is mimicked. We conclude that for realistic values of the baryon density, an SKA-like galaxy survey will not be able to detect this feature with standard correlation function analysis.
\end{abstract}

\maketitle

\tableofcontents

\section{Introduction}

One of the most important probes of the large scale structures (LSS) of the universe is the two-point function of galaxies. Measurements of the two-point function (2pF) have been reported by different collaborations in the past years~\cite{Jones:2009yz,2013AAS...22110602D,2015ApJS..219...12A,Hinton:2016atz,Ata:2017dya,2017ApJS..233...25A} and, in the future, upcoming redshift surveys will probe the LSS of the universe at deeper redshift and for larger volumes~\cite{2011arXiv1110.3193L,Maartens:2015mra}, with unprecedented precision. \\

With these surveys we can nicely link late-time measurements to early-time physics. The most striking example are the acoustic oscillations in the primordial plasma - first predicted in~\cite{Sunyaev:1970er,Sunyaev:1970eu,Peebles:1970ag} - which leave their imprint in cosmological observables. Their measurement is considered as one of the most important successes of the $\Lambda$CDM model. In the cosmic microwave background, the scale of the baryon acoustic oscillations (BAO) is a probe of the sound horizon at decoupling and it manifests itself as a series of peaks in the angular spectrum~\cite{Ade:2015xua}. A similar feature can also be seen in the matter power spectrum~\cite{Meiksin:1998ra}, while, in the 2pF, the same physics is responsible for a single peak located at a comoving distance slightly bigger - as we will explain in section~\ref{sec1} - than the sound horizon at decoupling. \\

The BAO peak in the correlation function has been first measured in~\cite{Eisenstein:2005su}. Since then it has been systematically used as a standard ruler to probe the distance-redshift relation~\cite{Beutler:2011hx,2011MNRAS.418.1707B,2012MNRAS.427.3435A}, in order to constrain the cosmic expansion history~\cite{Eisenstein:1998tu}. The peak is also sensitive to other cosmological parameters~\cite{vanDaalen:2015oul,Baumann:2017lmt,Baumann:2018qnt,Baumann:2017gkg}. A complication arises as the position of the peak measured with data cannot be fitted with linear theory: non-linearities affect both the position and the shape of the BAO feature~\cite{Eisenstein:2006nj,Crocce:2007dt,Smith:2007gi,McCullagh:2014jsa,Vlah:2015zda,Anselmi:2015dha}. \\

Here we consider a second feature: a trough in the correlation function positioned at twice the distance of the peak. The existence of this feature is implied by the well-known heuristic argument that is commonly used to explain the BAO peak (see section~\ref{sec1}), but rarely mentioned in the literature. In section~\ref{sec2} we mimic the fitting procedure - used by galaxy surveys to measure the peak position - to study the expected detection significance of the second feature for an SKA-like survey.

\section{The $2^\text{nd}$ feature}\label{sec1}
We outline in this section the heuristic argument given in the seminal paper~\cite{Eisenstein:2006nj} and summarized in the review~\cite{Bassett:2009mm}. This will give us insight on how this argument implies a second feature in the correlation function. The technical foundations can be found in~\cite{Bashinsky:2000uh,Bashinsky:2002vx,Montanari:2011nz}.\\

\begin{figure}[ht]
\includegraphics[scale=0.57]{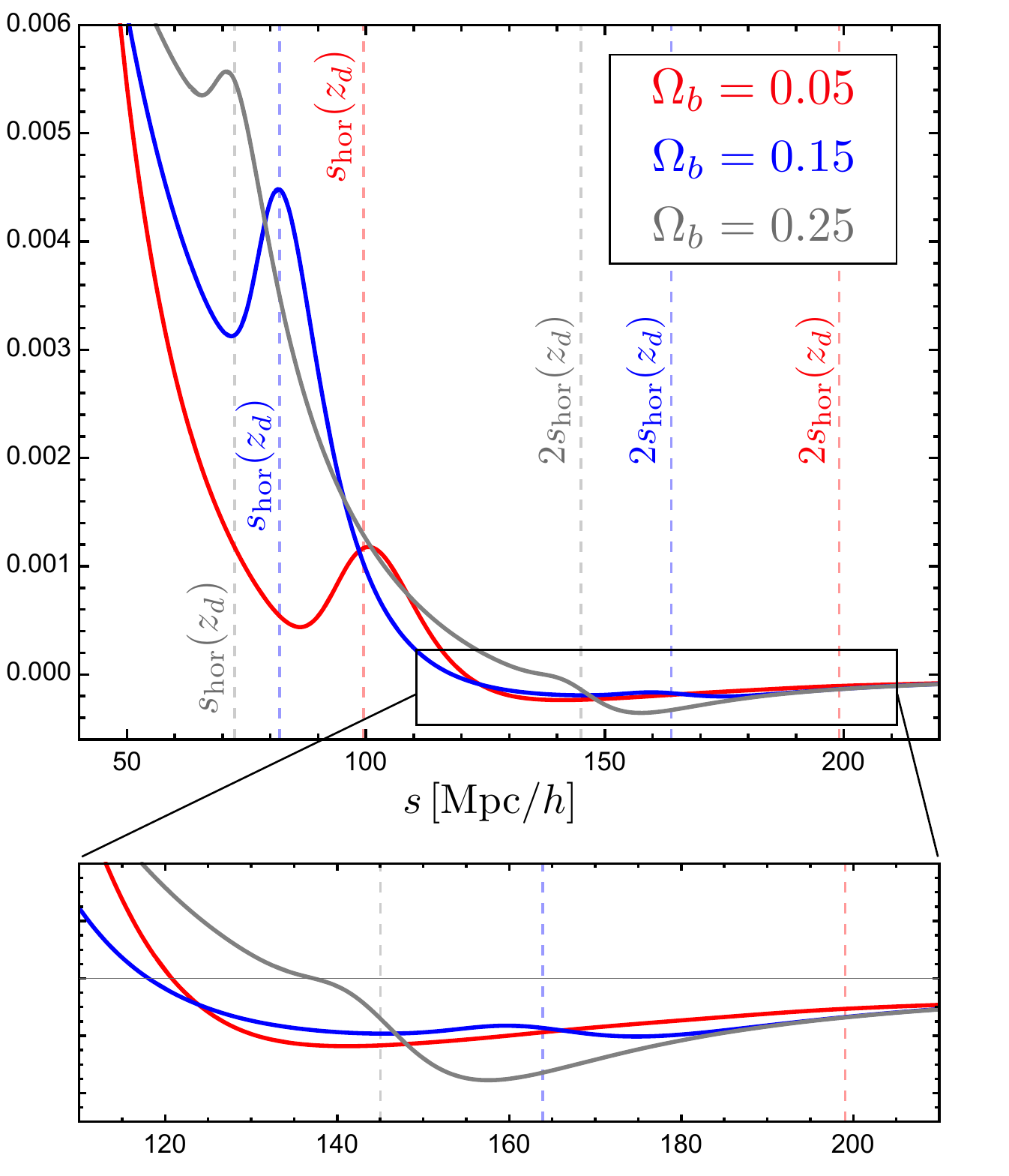}
\caption{\label{fig:allxi} The angle-averaged matter correlation function at $\bar z =1$ for three different cosmologies with $\Omega_b = 0.25$ (gray), $\Omega_b = 0.15$ (blue) and $\Omega_b = 0.05$ (red). The total matter density is fixed to $\Omega_m =0.31$ and the sound horizon at the drag epoch for the different models is shown.}
\end{figure}

Let us focus on some initial over-dense point in the primordial plasma - when baryons are tightly coupled to photons via Thomson scattering. If the fluctuations are adiabatic the over-density will be shared by all species: in particular a region over-dense in photons will also have an over-pressure with respect to its surroundings. This pressure imbalance causes an acoustic wave in the baryon-photon plasma which travels at the speed of sound $c_s$ until baryons decouple from the photons. When this happens the baryon's speed of sound goes to zero and the wave is frozen: the initial over-density is now composed only of dark matter while baryons have created an over-dense spherical region around the initial point. Every over-density will behave as we just described and the net result is that matter is more likely to cluster with a correlation length corresponding to the sound horizon at decoupling. It is clear that this process, as we have already anticipated, is responsible for the BAO peak: the correlation function is defined as the excess probability (over Poisson noise) of finding two tracers separated by a comoving distance $s$ and hence it peaks for $s\sim s_\text{hor}$, where the comoving sound horizon is defined as
\be
s_\text{hor} = \int\limits_0^{t_\text{drag}} dt\, c_s(t) (1+z) \,.
\label{shori}
\ee
and $t$ is cosmic time. The end of the Compton drag epoch $t_\text{drag}$ is the time at which the baryons are released from the drag of photons (at a later time than photon-decoupling, roughly at $z_\text{drag} \sim 1000$). It is the moment at which the baryons' velocity decouples from the photons. If we go back to our idealistic picture of a dark matter perturbation surrounded by a spherical shell of baryons we see that the correlation will not only be enhanced at $s_\text{hor}$ but, as all the baryons are in the shell, we will also get a trough when the correlation length reaches the diameter of the shell: $2 s_\text{hor}$. In other words - in our idealized picture - as long as the correlation length is $<2 s_\text{hor}$ the baryon-baryon correlation contributes to the matter 2pF as it is always likely to `find' two baryons in the shell. On the other hand, when the correlation length reaches the diameter of the shell, the baryon-baryon contribution has a sharp trough. This $2^\text{nd}$ feature (2FT) is illustrated in figure~\ref{fig:allxi}: we can clearly see the drop in the 2pF at  $2 s_\text{hor}$ for high values of $\Omega_b$, while the feature is less pronounced when baryons contribute less to the energy-density budget.

One might notice that, even in linear theory, the position of the acoustic peak is not exactly centered on $s_\text{hor}$: this is a known effect due to imperfect baryon-photon coupling, which allows photons to diffuse out of the perturbation and drag the baryons with them\footnote{In Fourier- and angular- space this effect is responsible for the \emph{Silk damping}.}, and velocity overshoot - both discussed in detail in~\cite{Sanchez:2008iw}. Note that the dark matter over-density does not remain at the center of the shell as it is gravitationally bound to the outgoing species, this does however not change the position of the peak. Despite these complications the peak is an extremely interesting cosmological observable as it is sensitive to a range of cosmological parameters. For example $s_\text{hor}$ is directly related to the sound speed $c_s$ via Eq.~(\ref{shori}) which, in turn, is related to the $\Omega_b$ and $\Omega_\gamma$ ratio. The positions of the features are also sensitive to the expansion history prior to decoupling as the propagation time of the sound wave depends on the expansion rate, introducing for example a subtle dependence on $\Omega_\nu$ (see~\cite{Thepsuriya:2014zda} for a comprehensive treatment). Finally, measuring the positions of the features as a function of redshift - using them as statistical standard rulers - constrains the late-time expansion rate and gives informations on $\Omega_m$, $\Omega_\Lambda$ and the equation of state of dark energy $w$. \\

We could naively think that, since we are searching for a feature at twice the separation of the BAO peak, we are safe from non-linear effects at these very large scales. This is only partially correct. Non-linear effects on the BAO peak come in two aspects: a broadening of the feature and a shift of the peak position. The damping effect is easily understood in real space, where non-linear physics can move the tracers around, on the scale of $\sim 10 \, \text{Mpc}$, pulling them out of the $100 \, \text{Mpc}/h$ shell and hence broadening the peak feature~\cite{McCullagh:2014jsa}. In Fourier space this effect is responsible for the smoothing of the subsequent peaks in the power spectrum (see e.g.~\cite{Anselmi:2012cn}). The fact that we are looking at two galaxies at a distance where linear physics should give an adequate description is not important in this case: the local non-linearities around the two tracers have an observable (and important) effect. For this reason we expect the 2FT to suffer from the same non-linear correction to its shape as it is not protected from non-linear broadening. We stick to the linear description of the 2pF in this work where the feature is sharper and therefore we will overestimate the detection significance in section~\ref{sec2}. This does not change our conclusions. On the other hand, in order to induce a shift in the position of the feature, non-linear physics has to coherently and systematically move tracers separated by $s_\text{hor}$ or $2 s_\text{hor}$ either closer or further away from each other. The small shift of the BAO peak has been widely investigated~\cite{Smith:2007gi,Desjacques:2010gz,Baldauf:2015xfa,Anselmi:2015dha,Obuljen:2016urm,Blas:2016sfa} and in this sense the fact the 2FT is located at larger scales means it will be less affected, as the position is only sensible to non-linear effects at the $\sim 200 \, \text{Mpc}/h$ scale.
\subsubsection{Fourier space}
\begin{figure}
\includegraphics[scale=0.55]{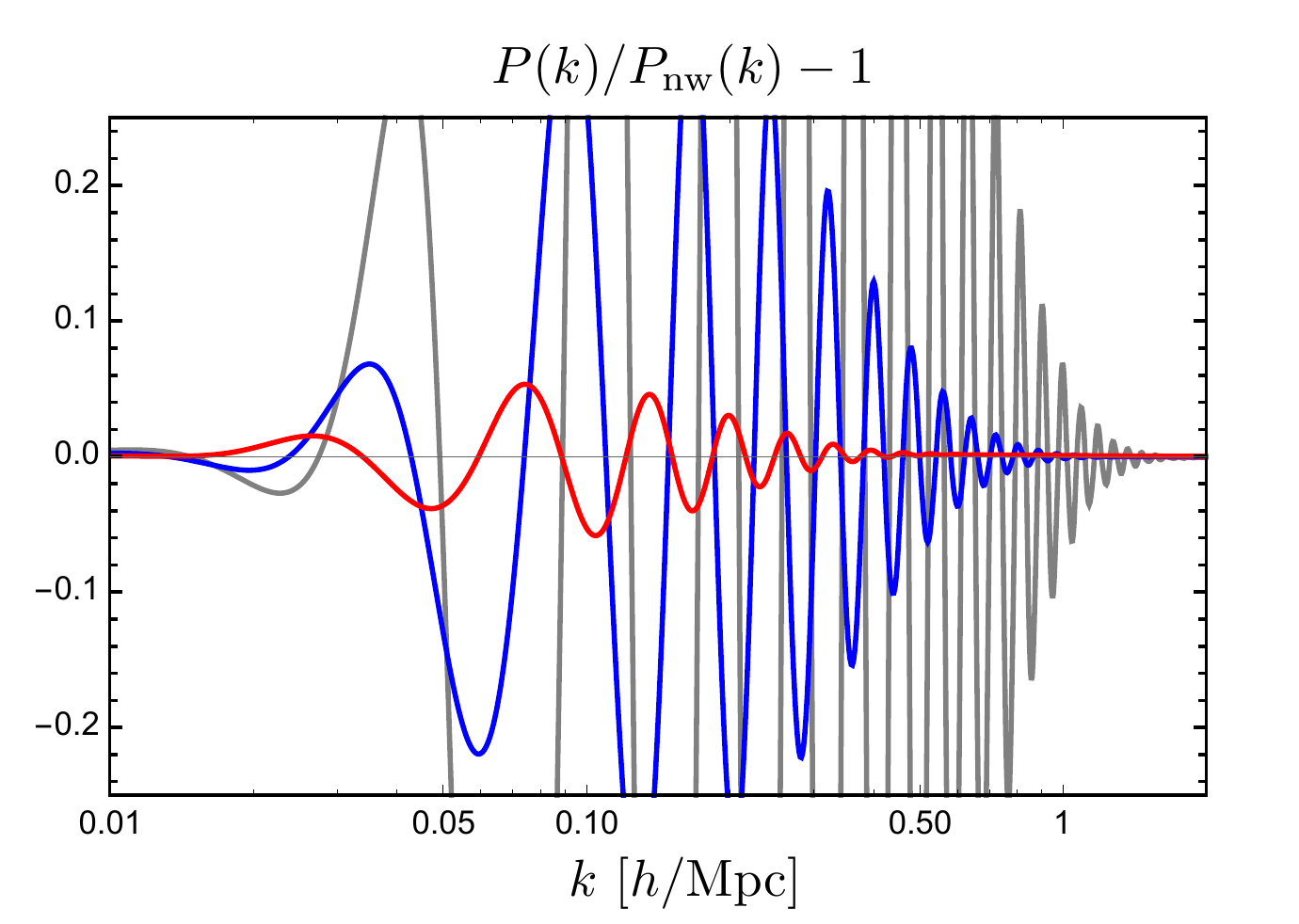}
\caption{\label{fig:deltapk} The residual BAO oscillations $P(k)/P_\text{nw}(k)-1$ once the broadband shape is factored out with the no-wiggle power spectrum $P_\text{nw}$, defined in Eq.~(\ref{nwpk}). Color coding as in fig.~\ref{fig:allxi}.}
\end{figure}

Let us now discuss how the simple picture depicted in this section is translated in Fourier space\footnote{This argument is discussed in \url{https://www.cfa.harvard.edu/~deisenst/acousticpeak/spherical_acoustic.pdf}: a short but vey nice essay which, to our knowledge, is the only place where the 2FT is briefly mentioned for the purely-baryonic case.}. The matter transfer function can be written as
\be
T_m(k) = f_\text{cdm} T_\text{cdm}(k) +  f_b T_b(k) \,,
\ee
where  $f_\text{cdm} = \Omega_\text{cdm}/\Omega_m$, $f_b = \Omega_b/\Omega_m$ and we drop the redshift-dependence here. $T_\text{cdm}$ is the smooth contribution of cold dark matter to the transfer function while $T_b \sim \sin( s_\text{hor} k )$ contains the `sine-wave' oscillations of the BAO. The power spectrum is then proportional to 
\be
k^3P(k) \sim f_\text{cdm}^2 T^2_\text{cdm} +2f_\text{cdm}  f_b T_\text{cdm} T_b +  f^2_b T^2_b \,.
\ee
For $f_b \ll f_\text{cdm}$ the last term is subdominant and $P(k)$ has the familiar shape of a superposition of a smooth function and a `sine-wave'. On the other hand for $f_b \gtrsim f_\text{cdm}$ the squared oscillations start to dominate, increasing both the frequency of the BAO and their amplitude  (see fig.~\ref{fig:deltapk}). When we Fourier transform to obtain the 2pF, the `sine-wave' part of $P(k)$ contributes to the BAO peak while the `sine-square' part from $T_b^2$ is responsible for the feature at twice the peak distance as it oscillates with twice the frequency.  Hence, also in Fourier space, bigger values of $f_b$ correspond to a more pronounced 2FT. 
\section{Fitting Methodology}\label{sec2}
We now want to gain insight on the ability of galaxy surveys to detect the second feature in the matter two-point function. The procedure to detect the BAO peak is now well-established (as described in~\cite{Xu:2012hg} and used e.g. in~\cite{Magana:2013wpa}) and we follow it here for the 2FT. The BAO detection is usually quoted as the $\chi^2$ difference between the best fit model and the model with no features. In other words we study how reliably we can reject a no-feature model. We do not deal with real data here but generate a fake `data' vector $\boldsymbol{\xi}$ from our fiducial model and compute the $\chi^2$ from the standard definition
\be
\chi_\text{fit} ^2(\alpha)= (\boldsymbol{\xi}_\text{fit}(\alpha) - \boldsymbol{\xi})^\text{T} \mathbf{C}^{-1} (\boldsymbol{\xi}_\text{fit}(\alpha) - \boldsymbol{\xi}) \,,
\label{chi2eq1}
\ee
where $\mathbf{C}$ is the covariance matrix for the fiducial model, defined in appendix~\ref{a:fcovmat}. The quantity $\alpha$ is the scale dilatation parameter  which measures the position of the feature (being the BAO peak or the 2FT) with respect to the fiducial model. In real data analyses $\alpha$ is a measure of 
\be
\alpha = \frac{D_V (z) s_{*,\text{fid}}}{D_V^\text{fid} (z) s_* }
\label{alpdef}
\ee
where the subscript `fid' means `fiducial',  $s_*$ is the comoving position of the feature and $D_V$ is the spherically averaged\footnote{Note that in real data situations both the angle averaged parameter $\alpha$ and an additional parameter $\varepsilon$ are considered: $\varepsilon$ parametrises the anisotropic clustering due to redshift-space distortions and due to an analysis where an incorrect cosmology is assumed.} distance defined as
\be
D_V(z) = \left[cz (1+z) D_A^2(z) \mathcal{H}(z)  \right]^{1/3} \,.
\ee
The parameter $\alpha$ is the measurement of the 2FT scale in the sense that it characterises any observed shift in the relative position of the acoustic feature in the data versus the model. The value of $\alpha$ which minimises $\chi^2$ is related to the feature position via Eq.~(\ref{alpdef}), and the feature position roughly marks $2 s_\text{hor}$.\\

In this work the fiducial model is generated starting from the linear matter power spectrum $P(k)$ obtained from {\sc class}~\cite{Blas:2011rf} (multiplied by the large-scale galaxy bias $b^2(z)$ given in appendix~\ref{a:fcovmat}) and converted into the full-sky correlation function~\cite{Tansella:2017rpi} using the {\sc coffe} \cite{Tansella:2018sld} code\footnote{Available at \url{https://github.com/JCGoran/coffe}.}. To mimic most BAO analyses we include the effect of redshift-space distortion in the 2pF but neglect other relativistic effects (such as lensing and the Doppler effect). We also neglect non-linear damping\footnote{We also do not consider a Gaussian damping term which is commonly introduced in the Fourier transform  $P(k) \rightarrow \xi(s)$ to improve numerical convergence as the {\sc coffe} code is based on the very reliable {\sc 2-fast} algorithm~\cite{Gebhardt:2017chz}.} and set the streaming scale to zero as our fiducial model - from which we draw data - is fully linear. To improve the signal-to-noise, galaxy surveys measure the spherically averaged two-point function $\xi_0(s)$ (the monopole) and the quadrupole $\xi_2(s)$ defined as
\be
\xi_\ell(s) = \frac{2\ell+1}{2} \int d\mu \, \xi(s,\mu) \mathcal{P}_\ell(\mu) \,,
\ee
where $\mathcal{P}_\ell$ is the Legendre polynomial of degree $\ell$ and $\mu$ is the orientation with respect to the line of sight at which we measure the 2pF. Our data vector is then given by
\be
\boldsymbol{\xi} = \begin{pmatrix} \boldsymbol{\xi}_0 \\ \boldsymbol{\xi}_2 \end{pmatrix} \,.
\ee
\begin{figure}[ht]
\includegraphics[scale=0.55]{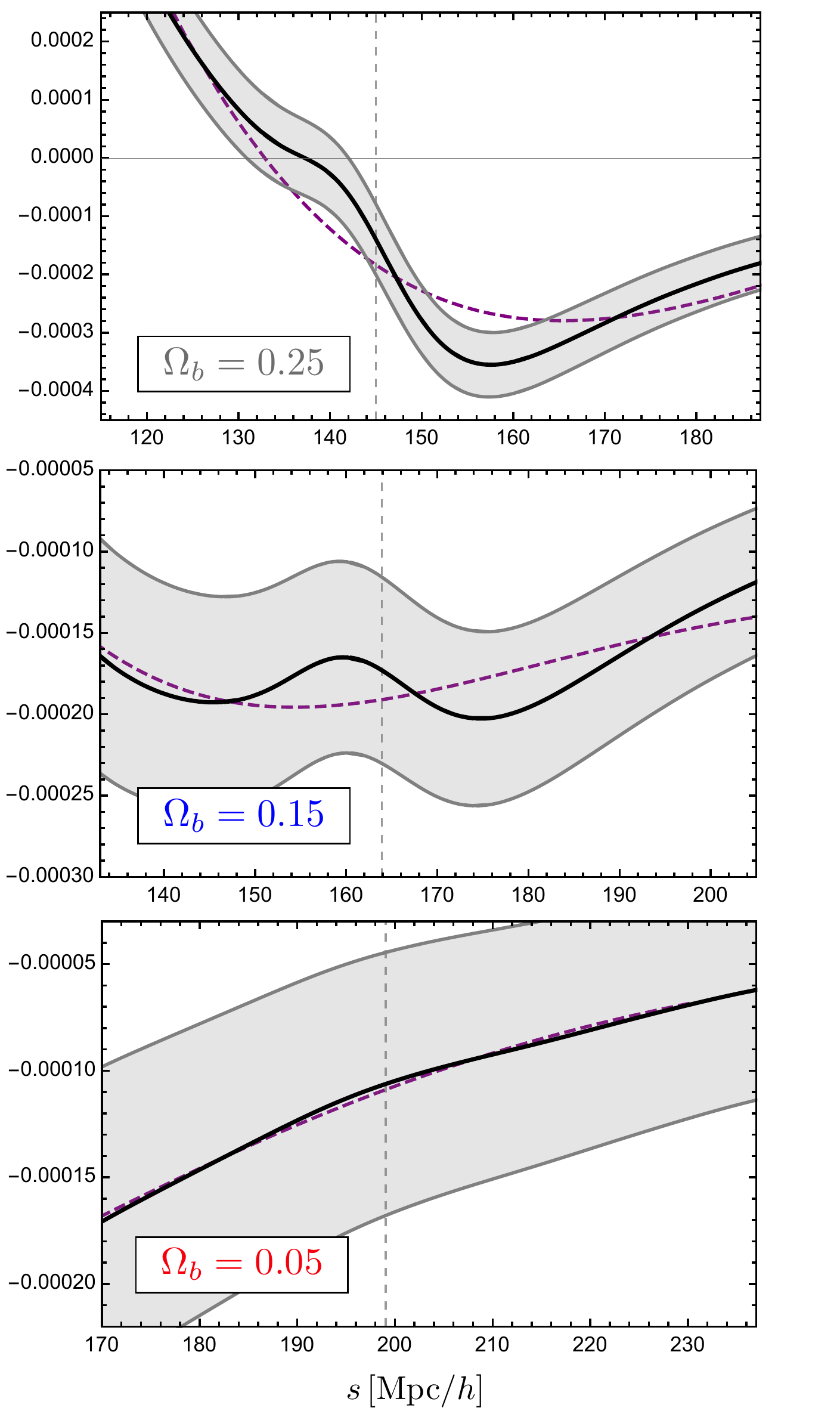}
\caption{\label{fig:zoomxi} A zoom of fig.~\ref{fig:allxi} for the range used in the $\Delta \chi^2$ estimation, where the $2^\text{nd}$ feature is clearly visible for high-baryon models. The gray region marks the error bars given by the covariance for an SKA-like survey and the purple dashed line is the best-fit non BAO (de-wiggled) model.}
\end{figure}
Fake `data' are generated for three different cosmologies: the fiducial \emph{Planck2015}\footnote{We set $h=0.676$, $\Omega_\text{cdm}=0.26$, $\Omega_b =0.048$, $\Omega_\Lambda = 0.68$. The primordial spectrum has $n_s = 0.96$ and $A_s = 2.22 \times 10^{-9}$ at $k_\text{pivot} = 0.05 \, \text{Mpc}^{-1}$ .} cosmology and two unrealistic toy models with $\Omega_b=0.25$ and $\Omega_b=0.15$ - keeping $\Omega_m$ and all the other parameters fixed - to illustrate the procedure in models where the 2FT is more pronounced. The binning of the data vectors ($\boldsymbol{\xi}_0$ and $\boldsymbol{\xi}_2$) is chosen in a range of $\sim 75 \, \text{Mpc}/h$ around the value $2 s_\text{hor}$ for each fiducial model and with a bin size $L_p = 3\, \text{Mpc}/h$, for 25 bins in total. The covariance matrix is computed for an SKA-like survey, with parameters given in appendix~\ref{a:fcovmat}.
\begin{figure}[ht]
\includegraphics[scale=0.55]{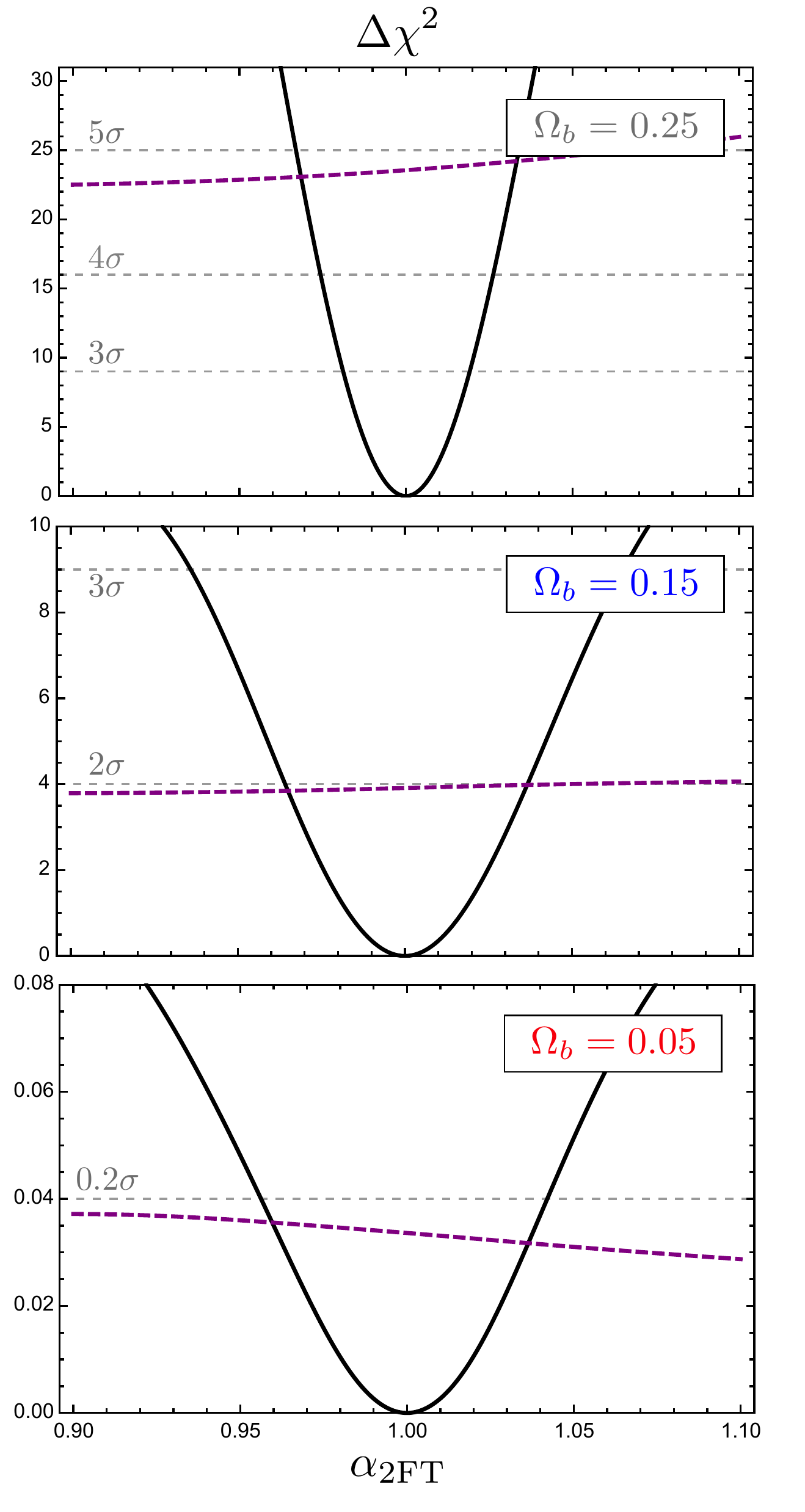}
\caption{\label{fig:chi2tot} The $\Delta \chi^2 (\alpha) = \chi^2(\alpha) -\chi^2_\text{min}$ as a function of the BAO-parameter $\alpha_{2FT}$. The solid line is the model of Eq.~(\ref{mod1}) while the dashed curve displays the same information for a no BAO model (Eq.~(\ref{mod2})), where $\Delta \chi^2$ is determined by subtracting the minimum $\chi^2_\text{min}$ from the BAO model.}
\end{figure}
\subsubsection{Fitting models}\label{s21}
To fit the correlation function we adopt two template models, one with the BAO peak and the 2FT and the other with no baryonic features. The fit is performed, as in recent BAO data analyses, with 5 parameters: a multiplicative bias $B$, the scale dilatation $\alpha$ and (as we are only interested in the position of the 2FT) a $2^\text{nd}$ order polynomial to marginalise over the broad-band shape of the multipoles. We then write
\be
\begin{split}
\xi_0^\text{fit}(s) &= B^2 \xi_0^\text{mod} (\alpha, s) + A_0(s) \,, \\
\xi_2^\text{fit}(s) &=  \xi_2^\text{mod} (\alpha, s) + A_2(s) \,,
\end{split}
\ee
where we define
\be
A_\ell (s) = \frac{a_{1,\ell}}{s^2} +\frac{a_{2,\ell}}{s} + a_{\ell,3} \quad ; \,\, \ell=0,2 \,,
\label{ploynu}
\ee
with three nuisance parameters per multipole $(a_1,a_2,a_3)$, to account for the overall unknown shape of the correlation function. A difference with the standard approach is that we set here $B=1$ for two reasons. Firstly, as we are not dealing with real data, we have full control on the linear bias parameter when we generate fake data from our fiducial model. Secondly we are not comparing different cosmologies (for which the amplitude of the feature might change) but the same cosmology with and without the feature. This also prevents the data to be fitted only by the quadratic polynomial $A_\ell$.\\

The first template model is simply given by
\be
 \xi_\ell^\text{mod}(\alpha,s) =  \xi_\ell^\text{fid} (\alpha s) \,.
 \label{mod1}
\ee
Note that when performing the BAO analysis in real space, it is standard practice to shift the all model as in Eq.~(\ref{mod1}). A different approach is usually employed in the Fourier space analysis where only the BAO oscillations are shifted.  As the nuisance parameters $a_{i,\ell}$ are marginalizing over the broadband shape of the multipoles, this has no effect~\cite{Anderson:2013zyy}. \\

The second template is the \emph{de-wiggled} model. It is a phenomenological prescription widely used in BAO analysis: it consists in generating a correlation function starting from a power spectrum $P_\text{nw}(k)$ in which the BAO features have been erased. To obtain $P_\text{nw}(k)$ we start with the Eisenstein\&Hu~\cite{Eisenstein:1997jh,Eisenstein:1997ik} approximated power spectrum $P_\text{EH}(k)$ and perform a Gaussian smoothing on the ratio $P(k)/P_\text{EH}(k)$:
\be
P_\text{nw}(k) = P_\text{EH}(k) \,\mathcal{S} [P(k)/P_\text{EH}(k)] \,,
\ee
where $\mathcal{S}$ schematically represents the smoothing. The no-wiggle spectrum is then given by~\cite{Vlah:2015zda}
\be
\begin{split}
\frac{P_\text{nw}(10^{k_{log}})}{P_\text{EH}(10^{k_{log}})} &= \frac{1}{\sqrt{2\pi}\lambda} \int d q_{log} \bigg[ \frac{P(10^{q_{log}})}{P_\text{EH} (10^{q_{log}})} \\
&\times  \text{Exp} \left(-\frac{1}{2\lambda^2}(k_{log}-q_{log})^2 \right) \bigg] \,,
\end{split}
\label{nwpk}
\ee
where $\lambda$ is a parameter that controls the size of the smoothing. We found the best results for $\lambda = 0.14 \, \text{Mpc}/h$. In figure~\ref{fig:deltapk} we plot the fractional difference of the no-wiggle power spectrum and the linear one. The multipoles of the correlation function with no feature $\xi^\text{nw}_\ell(s)$ are then generated by feeding {\sc coffe} with $P_\text{nw}$ and the second template model is given by
\be
 \xi_\ell^\text{mod}(\alpha,s) =  \xi_\ell^\text{nw} (\alpha s) \,.
  \label{mod2}
\ee

For every value of $\alpha$ we fit the remaining parameters to minimise the $\chi^2$ for both models. We chose only one fiducial redshift $z=1$, hence we require the size of the redshifts bin of the survey to be $\Delta z \gtrsim 0.2$. We focus here only on one redshift bin as the shape of the correlation function is nearly constant at large scales for the depth accessible by galaxy surveys and the analysis is trivially extended to more bins, given also the fact that we can treat them as uncorrelated to a good approximation. In figure~\ref{fig:zoomxi} we show the fiducial model monopole $\xi_0^\text{fid}(s)$ with the error bars obtained from Eq.~(\ref{covxi}), together with the best fit no-feature model of Eq.~(\ref{mod2}). Clearly as $\Omega_b$ decreases the feature is less and less pronounced and the no-feature model is an increasingly better fit to the data. For a realistic model with $\Omega_b \simeq 0.05$ the 2FT is barely visible and lies completely within the error bars. This situation is reflected when we compare the $\Delta \chi^2(\alpha) = \chi^2(\alpha) -\chi^2_\text{min}$ for the two templates.  We can read off the detection significance for the 2FT in fig.~\ref{fig:chi2tot}. In the two toy models - with an unrealistically high baryon fraction - the no-feature templates are disfavored at $\sim5\sigma$ and $\sim 2\sigma$ respectively. The realistic model \emph{Planck2015} shows no preference for the template which correctly describes the 2FT compared to the smoothed template. We have checked that these results marginally change when we vary the order of the polynomial fit in Eq.~(\ref{ploynu}). Note that the no-BAO model has a broad $\chi^2$ as the lack of features makes the scale less constrainable.

\section{Conclusions}\label{s:con}
In this work we have introduced a second feature in the matter correlation function. This feature, positioned at twice the distance of the BAO peak, is understood - in the early universe - as a trough in the baryon-baryon correlation for separations bigger than twice the sound horizon at $t_\text{drag}$. The feature is clearly visible in models with an high baryon fraction but in a realistic cosmological model it is a very small effect. We proved this with a $\chi^2$ analysis that showed how - in an SKA-like survey - it is not possible to distinguish between the models with and without the feature. We have considered only one redshift bin and it is possible to increase the detection significance by a factor $\sim \sqrt{N}$ by considering $N$ bins; however, the analysis requires $\Delta z_\text{bin} \gtrsim 0.2$ hence limiting the number of bins $N$ in which we can split a galaxy catalog. Furthermore, the fact that at $2 s_\text{hor}$ the error is cosmic variance dominated suggests that the two-point function is not the best observable to detect this feature. \\

In a Fourier space analysis the effect described here is correctly modeled if the template $P(k)$ is generated from a Boltzmann code such as {\sc class}~\cite{Blas:2011rf,DiDio:2013bqa} or {\sc camb}~\cite{Lewis:1999bs}.\\

It is nevertheless interesting to study if other observable - e.g. intensity mapping - are more sensitive to this feature which, if detected, would provide an additional probe for early-time cosmology. We leave this matter for future work.

\acknowledgments
We wish to thank Jonathan Blazek, Enea Di Dio, Ruth Durrer, Fabien Lacasa, Martin Kunz and Elena Sellentin for useful discussions. We additionally acknowledge useful comments on the draft of this paper from Ruth Durrer, Martin Kunz and Davide Racco. The author acknowledges support by the Swiss National Science Foundation.
\newpage


\appendix

\section{Covariance matrix}\label{a:fcovmat}

The simplest estimator we can construct for the multipoles is given in~\cite{Tansella:2018sld} as
\be
\hat \xi_\ell (x) =\frac{2\ell+1}{4\pi} \frac{L_p^5}{x^2 V} \sum_{ij} \De_i \De_j \PP_\ell(\mu_{ij}) \delta_K(x_{ij}-x) \,, \nonumber
\ee
and the covariance matrix is defined by
\bea
&\text{cov}_{\ell\ell'}^{(\xi)}(x_i,x_j) \equiv  \Big\langle \hat \xi_\ell (x_i) \hat \xi_{\ell'}(x_j) \Big\rangle- \Big\langle  \hat \xi_\ell (x_i)\Big\rangle \Big\langle\hat \xi_{\ell'}(x_j) \Big\rangle \,. \nonumber
\eea
The variance of the number counts has two contributions
\be
\langle \De_i\De_j\rangle = \frac{1}{d\bar N} \de_{ij} + C_{ij} \,.
\ee
The first term, where $d\bar N$ is the average number of tracers per pixel, accounts for shot noise, coming from the fact that we are Poisson sampling from the underlying density distribution and it is a contribution to the correlation at zero separation. The second term is the cosmic variance contribution. For simplicity we perform the covariance calculation in the flat-sky approximation where the 2pF is simply written 
\be
\begin{split}
&\xi(\bar z, s, \mu)_\text{flat-sky} = \frac{D^2_1(\bar z)}{D_1^2(0)} \bigg[c_0(\bar z) I^0_0(s) \\&- c_2(\bar z) I^0_2(s) \PP_2(\mu) +c_4(\bar z) I^0_4(s) \PP_4(\mu)\bigg] \,,
\end{split}
\label{fsxi}
\ee 
with
\begin{align}
c_0 &=  b^2+\frac{2}{3}bf +\frac{f^2}{5} \,,\\
c_2 &=  \frac{4}{3} bf +\frac{4}{7} f^2 \,, \\
c_4 &= \frac{8}{35} f^2\,.
\end{align}
Here $D_1(\bar z)$ is the growth factor,
\be\label{e:growth}
f(\bar z) =\frac{d\ln D_1}{d\ln(a)} 
\ee
is the growth rate and we have also introduced the integrals
\be
I_\ell^n(s) = \int \frac{dk\, k^2}{2 \pi^2} \,P(k)|_{z=0} \,\frac{ j_\ell( k s )}{(k s)^n} \,.
\ee
Assuming gaussianity (i.e. we write 4-points functions as products of 2-points functions) we follow the procedure outlined in~\cite{Tansella:2018sld,Bonvin:2017req} and obtain: 
\be
\begin{split}
&\text{cov}_{\ell\ell'}^{(\xi)}(x_i,x_j) = \frac{ i^{\ell-\ell'} }{V} \Bigg[  \frac{2\ell+1}{2\pi \bar n^2 L_p x_i^2} \de_{ij} \de_{\ell\ell'} \\&+ \frac{1}{\bar n} \mathcal{G}_{\ell\ell'} (x_i,x_j) \sum\limits_\sigma c_\sigma\tj{\ell}{\ell'}{\sigma}{0}{0}{0}^2\\
&+ \mathcal{D}_{\ell\ell'} (x_i,x_j) \sum\limits_\sigma \tilde c_\sigma\tj{\ell}{\ell'}{\sigma}{0}{0}{0}^2 \Bigg] \,,
\end{split}
\label{covxi}
\ee
where $\bar n$ is the mean number density in the redshift bin and we have defined
\begin{align}
&\mathcal{G}_{\ell\ell'} (x_i,x_j) = 2 w(\ell,\ell')\! \!\int dk \, k^2  P(k,\bar z) j_\ell( k x_i) j_{\ell'}(k x_j) \,, \nonumber \\
& \mathcal{D}_{\ell\ell'} (x_i,x_j) = w(\ell,\ell') \!\!\int dk \, k^2  P^2(k,\bar z) j_\ell( k x_i) j_{\ell'}(k x_j) \,, \nonumber 
\end{align}
with
\be
w(\ell,\ell') =  \frac{(2\ell+1)(2\ell'+1)}{\pi^2} \,,
\ee
and the modified redshift-distortion coefficients
\bea
&&\tilde  c_0 = c_0^2 +\frac{c_2^2}{5}+\frac{c_4^2}{9} \,,\\
&&\tilde  c_2 = \frac{2}{7}c_2 (7c_0+c_2) +\frac{4}{7} c_2 c_4+\frac{100}{693}c_4^2 \,, \\
&&\tilde  c_4 = \frac{18}{35} c_2^2 +2 c_0 c_4 +\frac{40}{77} c_2 c_4 +\frac{162}{1001} c_4^2 \,,\\
&&\tilde  c_6 = \frac{10}{99} c_4 (9 c_2 +2 c_4) \,,\\
&&\tilde  c_8 = \frac{490}{1287} c_4^2 \,.
\eea
The covariance matrix $\mathbf{C}$ in Eq.~(\ref{chi2eq1}) is then written as
\be
\mathbf{C} = \begin{pmatrix}
   \text{\bf{cov}}_{00} &  \text{\bf{cov}}_{02}  \\
    \text{\bf{cov}}_{20} &  \text{\bf{cov}}_{22}  
\end{pmatrix} \,,
\ee
where $\text{\bf{cov}}_{\ell\ell',ij} = \text{cov}_{\ell\ell'}^{(\xi)}(x_i,x_j)$.\\

For an SKA-like survey the parameters are taken from~\cite{Bull:2015lja} (Table 3). We consider a single redshift-bin centred at $\bar z=1$ with thickness $\Delta z = 0.2$, sky-coverage $f_\text{sky} \simeq 0.72 $, mean number density $\bar n \simeq 8.7 \times 10^{-4} \, \text{Mpc}^{-3}$ and bias $b \simeq 1.3$.

\vspace{1cm}
\newpage
\bibliographystyle{utcaps}
\bibliography{2nd-refs}

\end{document}